\documentclass
[aps,prl,twocolumn,floatfix,english,showpacs,10pt]{revtex4-2}%
\usepackage{graphicx}
\usepackage{booktabs}
\usepackage{amsmath}
\usepackage{physics}
\usepackage{amssymb}
\usepackage{colordvi}
\usepackage{verbatim}
\usepackage{xcolor}
\usepackage{mathrsfs}
\usepackage{epsfig}
\usepackage{lipsum}
\usepackage{amsfonts}
\usepackage{makecell}
\usepackage{esint}
\usepackage[unicode=true, breaklinks=false, pdfborder={0 0 1}, backref=false,
colorlinks=true, linkcolor=blue, urlcolor=blue, citecolor=blue]{hyperref}%
\providecommand{\U}[1]{\protect\rule{.1in}{.1in}}
\providecommand{\U}[1]{\protect\rule{.1in}{.1in}}
\setcitestyle{numbers,square}
\begin{document}

\title{Spin Radiation of Electrons, Excitons, and Phonons}
\author{Chengyuan Cai}
\affiliation{School of Physics, Huazhong University of Science and Technology, Wuhan 430074, China}

\author{Tao Yu}
\email{taoyuphy@hust.edu.cn}
\affiliation{School of Physics, Huazhong University of Science and Technology, Wuhan 430074, China}

\date{\today}

\begin{abstract}
In the celebrated Stern-Gerlach experiment an inhomogeneous static magnetic field separates a beam of \textit{charge-neutral} atoms with opposite spins, thereby driving a ``spin current" normal to the propagation direction. Here we generalize it to the dynamic scenario by demonstrating a spin transfer from an AC \textit{inhomogeneous} magnetic field to electrons or charge-neutral excitons and phonons.  We predict that parametric pumping can efficiently radiate the angular momentum of local AC magnetic sources to their DC spin currents with van der Waals semiconductors as prototypes. This mechanism brings a unified and efficient paradigm in the spin transport of distinct mobile carriers.

\end{abstract}
\maketitle

\textit{Introduction}.---The flow of electron spins or spin current is a fundamental physical concept that plays an important role in understanding the conversion between angular momentum in different disguises~\cite{Lenk,Maekawa_2023,chirality,orbital_transport,Fert}. Besides electrons,
angular momentum and magnetic moments can also be carried by bosonic information carriers such as magnons~\cite{Chumak,Brataas,Non_Hermitian,Barman}, photons~\cite{Bliokh,Lloyd}, 
chiral phonons~\cite{phonon_spin_1,phonon_spin_2,phonon_spin_3}, as well as
excitons~\cite{High,valley_Zeeman_1,valley_Zeeman_2,valley_Zeeman_3}. Spin pumping by magnetic
contacts~\cite{Tserkovnyak,Tserkovnyak2} and spin-Hall effect~\cite{Sinova,Hirsch,Jungwirth} are popular approaches generating electron spin currents in metals but are less appealing
for semiconductors and low-dimensional van der Waals materials due to Schottky barriers and
electronic structure mismatch~\cite{Avsar}. Interband photon absorption of specific optical frequencies may create spin polarization of electrons in the presence of optical selection rules~\cite{Sipe,HongxingXu,Aronov,Edelstein,Ivchenko,MoS2_RMP}, but usually not a spin current.  Its spin current may be driven when specific spin-orbit coupling or magnetism exists by, e.g., the photo-galvanic effect~\cite{Ivchenko,Belinicher,Asnin,JunChen,XixiTao,HaoJin}, spin-galvanic effect~\cite{Ivchenko_spin,Ganichev}, optical intersite spin transfer effect~\cite{Elliott,Dewhurst,Tengdin,Willems}, or photo-spin-voltaic effect~\cite{Ellsworth,Li,HaoweiXu}. Still, these effects request specific materials/heterostructures or specific light frequencies. These approaches are difficult to apply to charge-neutral carriers such as phonons and excitons, which have become important information carriers utilized in modern quantum nanodevices~\cite{phonon_spin_1,phonon_spin_2,phonon_spin_3,High,valley_Zeeman_1,valley_Zeeman_2,valley_Zeeman_3}.

Generation of spin current of charge-neutral carriers may be historically traced to the celebrated Stern-Gerlach experiment~\cite{Stern_Gerlach}, where the gradient of static magnetic fields separates a beam of charge-neutral atoms with opposite spins, thereby driving a spin current normal to the propagation direction.    
Oscillating magnetic fields may act as a periodic driving force for spin oscillation in antiferromagnets~\cite{Galkin,Satoh,Kampfrath}, charge-neutral 
excitons~\cite{CrSBr_1,CrSBr_2,CrSBr_3} and chiral phonons~\cite{YafeiRen}, as well as electrons, which can transfer energy, angular momentum, and linear momentum to the carriers. However, in the intraband optical transition, plane-wave optical photons hold very little momentum and cannot directly generate a spin current for charge-neutral carriers and electrons in extended materials since energy and momentum conservations cannot be simultaneously satisfied.  Localized AC magnetic fields are immune to momentum conservation, which can provide local forces exerted on the carrier spins analogous to the Stern-Gerlach experiment~\cite{Stern_Gerlach}. Still, this force is periodic rather than static. We thereby raise the question and investigate whether a DC spin current of distinct mobile carriers can be pumped by the periodic local force on spins as a ``dynamical" analog of the Stern-Gerlach experiment.

\begin{figure}[pth]
\centering
\includegraphics[width=0.46\textwidth]{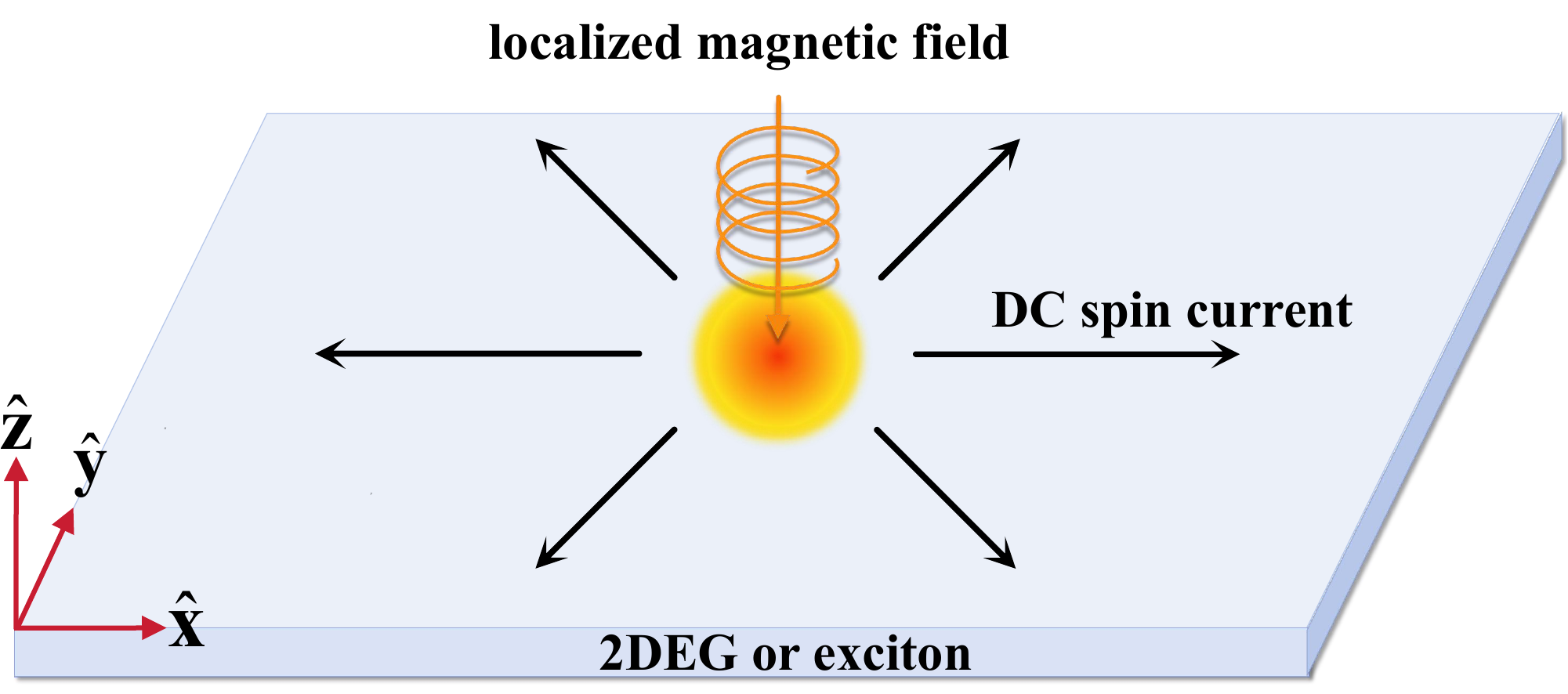}\caption{Radiation of DC spin
current of electrons or excitons when pumped by a focused magnetic (optical or microwave) field with circular polarization.}
\label{model}
\end{figure}

In this Letter, we predict \textit{intraband} angular momentum transfer between a
focused radio-frequency (rf) or terahertz (THz) radiation and the electrons or charge neutral carriers in conductors, semiconductors, and van der Waals materials, which are very different from the creation of interband electron-hole pairs by the polarized
\textit{electric} fields of THz/infrared radiation~\cite{Ivchenko,Belinicher,Asnin,Ivchenko_spin,Ganichev,XixiTao,WenYuShan}. Strongly localized near
fields may be generated by, \textit{e.g}., proximity excited nanomagnets~\cite{HanchenWang,2D_nonhermitian,Kamenetskii,Baumgaertl,evanescent_pumping},  focused laser beams, metallic nanostructures~\cite{nano_optics,plasmonics} or a scanning near-field optical microscope (SNOM)~\cite{Betzig,LeWang,HaominWang,Vincent}. We predict a parametric pumping mechanism that efficiently generates DC spin currents
carried by electrons, charge-neutral excitons, and phonons (Fig.~\ref{model}). Since the spin current is radiated from the local source, we term this phenomenon as  ``spin radiation" for short.

\textit{Inelastic spin-flip by photons}.---We first sketch the key physical processes by electrons and estimate the magnitude of DC spin currents
emitted by \textit{localized} AC magnetic (microwave or optical) fields in a setup as in
Fig.~\ref{model}. We consider a monochromatic magnetic field
$\mathbf{h}({\pmb\rho},t)=\sum_{\mathbf{q}}\left(  \mathbf{h}^{(+)}%
(\mathbf{q})e^{-i\omega t}+\mathbf{h}^{(-)}(\mathbf{q})e^{i\omega t}\right)
e^{i\mathbf{q}\cdot{\pmb\rho}}$~\cite{footnote} with frequency $\omega$
applied to a two-dimensional electron gas (2DEG) with an in-plane position vector
${\pmb\rho}=x\hat{\mathbf{x}}
+y\hat{\mathbf{y}}$. The Fourier components of a strongly localized field at the origin with polarization or \textquotedblleft
spin\textquotedblright\ along the out-of-plane $\hat{\mathbf{z}}$-direction read
\begin{equation}
\mathbf{h}^{(\pm)}(\mathbf{q})\approx(h_{0}/S)(1,\pm i,0)^{T},
\label{polarizedfield}
\end{equation}
where $h_{0}$ is the amplitude, $S$ is the sample area, and ``$+$" (``$-$")
corresponds to the positive (negative) circular polarization. It is wave number independent, so ${\bf q}$ in \eqref{polarizedfield} only indicates ``Fourier component". It couples with electrons by the Zeeman interaction $\hat{V}(t)=\mu_{0}\gamma_{e}\int%
\hat{\mathbf{s}}(\boldsymbol{\rho})\cdot\mathbf{h}(\boldsymbol{\rho
},t)d{\pmb\rho}$, in which $\mu_{0}$ is the vacuum permeability and
$\gamma_{e}$ is the effective gyromagnetic ratio of electrons, and excites a
non-equilibrium spin accumulation in the conduction band, which is the expectation value of the spin density operator
$\hat{\mathbf{s}}({\pmb\rho})=\sum_{\eta,\epsilon}(\hbar/2){\pmb\sigma}
_{\eta\epsilon}|{\pmb\rho},\eta\rangle\langle{\pmb\rho},\epsilon|$,
where ${\pmb\sigma}$ are Pauli matrices, $\{\eta,\epsilon\}=\{\uparrow,\downarrow\}$ denote electron spins along the $\hat{\mathbf{z}}
$-direction, and $|{\pmb\rho},\eta\rangle$ is an electron ket. \textcolor{blue}{Including the electric field ${\bf E}(\pmb{\rho},t)=-\partial{\bf A}(\pmb{\rho},t)/\partial t$ via the vector potential ${\bf A}(\pmb{\rho},t)$, the electron Hamiltonian
\begin{align}
 \hat{H}_e&=\int d {\pmb \rho}\left(\frac{\left(\hat{\bf p}-e{{\bf A}({\pmb\rho},t)}\right)^2}{2m^*}+\mu_0 \gamma_e \hat{\mathbf{s}}(\boldsymbol{\rho}) \cdot \mathbf{h}(\boldsymbol{\rho},t)\right)\nonumber\\
 &=\sum_{\bf k}\sum_{\eta=\{\uparrow,\downarrow\}}\left(\varepsilon_{\bf k}-\mu\right)|{\bf k},\eta\rangle\langle{\bf k},\eta|\nonumber\\
    &+\sum_{\zeta=\{\pm\}}\sum_{{\bf k},{\bf k}'}\sum_{\eta,\epsilon=\{\uparrow,\downarrow\}}{\cal G}^{(\zeta)}_{{\bf k}',{\bf k}}|_{\eta\epsilon}e^{-i\zeta\omega t}|{\bf k}',\eta\rangle\langle{\bf k},\epsilon|,
 \label{Hamiltonian}
\end{align}
where $\mathbf{k}=k_{x}\hat{\mathbf{x}}+k_{y}%
\hat{\mathbf{y}}$ is the electron wave vector, $m^*$ is the effective mass of electrons, $\mu$ is the chemical potential, and the coupling matrix ${\cal G}^{(\zeta=\pm)}_{\bf k',k}=\left[\frac{i\zeta e\hbar}{2m^*\omega}({\bf k+k'})\cdot{\bf E}^{(\zeta)}({\bf k'-k})\right]+({\hbar}/{2})\mu_0\gamma_e\pmb{\sigma}\cdot{\bf h}^{(\zeta)}({\bf k'-k})$ 
is contributed by the electric field that is independent of spins and Zeeman coupling.}

\textcolor{blue}{The spin injection rate (refer to the Supplemental Material (SM)~\cite{supplement} for details) 
\begin{widetext}
\begin{align}
\frac{\partial{\bf s}}{\partial t}\Big|_{\mathrm{DC}} 
&=\sum_{\mathbf{k},\mathbf{q}}\left[i\pi\left(\frac{\hbar\mu_0\gamma_e}{2}\right)^2\left(f_{\mathbf{q}
}-f_{\mathbf{k}}\right) \left(  \mathbf{h}_{\mathbf{q-k}}^{(+)\ast}\times\mathbf{h}
_{\mathbf{q-k}}^{(+)}\right)
\delta\left(  \varepsilon_{\mathbf{k}}+
\hbar\omega-\varepsilon_{\mathbf{q}}\right)+{\rm H.c.}\right]\nonumber\\
&+\sum_{\zeta=\pm}\sum_{\bf k,q}\frac{i\zeta\pi\hbar^2e\mu_0\gamma_e}{4m^*\omega}(f_{\bf q}-f_{\bf k})\left[\left(({\bf k+q})\cdot{\bf E}^{(\zeta)}_{\bf q-k}\right){\bf h}^{(\zeta)*}_{\bf q-k}-\left(({\bf k+q})\cdot{\bf E}^{(\zeta)*}_{\bf q-k}\right){\bf h}^{(\zeta)}_{\bf q-k}\right]\delta\left(  \varepsilon_{\mathbf{k}}+\zeta
\hbar\omega-\varepsilon_{\mathbf{q}}\right)
\label{spin_injection_rate}
\end{align}
\end{widetext}
contains contributions by electric and magnetic fields, noting that the $E^2$-term is traceless over the Pauli matrices and thereby has no contribution to the spin injection, 
where $f_{\mathbf{k}}=1/(e^{(\varepsilon
_{\mathbf{k}}-\mu)/(k_{B}T)}+1)$ is the
Fermi-Dirac distribution at temperature $T$. We demonstrate in the SM~\cite{supplement} that the second term of Eq.~\eqref{spin_injection_rate} is zero after summation over $\zeta=\pm$ since the Fourier components ${\bf h}^{(\zeta)*}_{\bf q}={\bf h}^{(-\zeta)}_{-\bf q}$ and ${\bf E}^{(\zeta)*}_{\bf q}={\bf E}^{(-\zeta)}_{-\bf q}$. 
This may be understood from the angular momentum conservation since the photon spin does not couple directly to the electric field.  Without significant spin-orbit coupling or optical selection rules, only the AC magnetic field contributes to the spin pumping.
} Nevertheless, the electric field may drive and heat the electrons, affecting the electron distribution function. This effect relies on the frequencies and sources, i.e., the rf stray fields of the ferromagnetic resonance of nanomagnets~\cite{Baumgaertl,evanescent_pumping,HanchenWang,2D_nonhermitian,Kamenetskii} or the THz optical field~\cite{nano_optics,plasmonics,Betzig,LeWang,Vincent,HaominWang}. The rf electric field by magnetization dynamics is a secondary effect and can be negligible; heating due to the THz field can be insignificant for 2DEG due to small conductivities~\cite{supplement,Asmar}.

\begin{figure}[pth]
\centering
\includegraphics[width=0.42\textwidth]{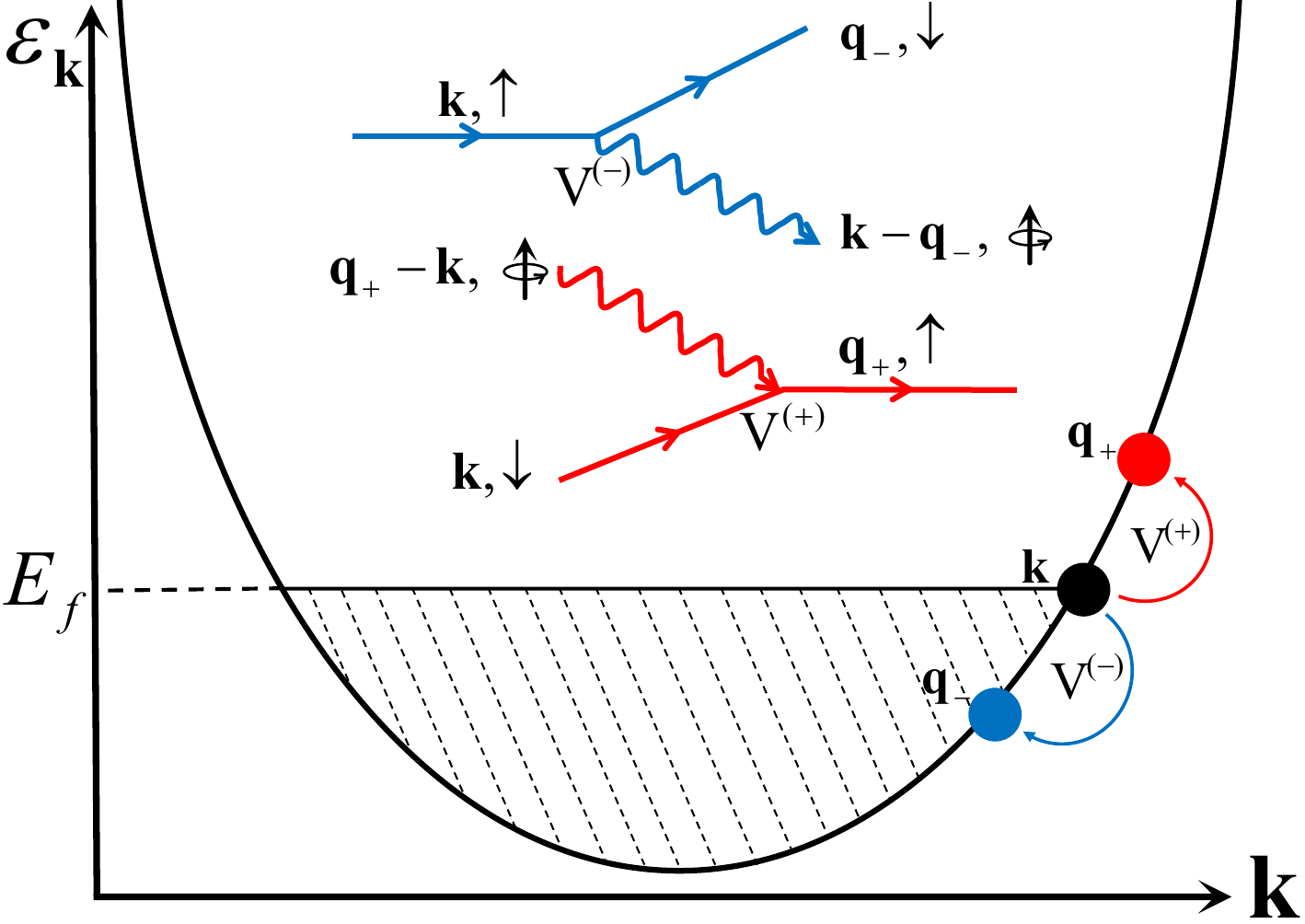}
\caption{Spin-transfer process in an electron gas induced by a circularly polarized photon. An electron at the Fermi energy can absorb a photon with energy to be excited to a higher energy state or emit a photon to a lower energy state under energy and linear/angular momentum conservation. Here state \textbf{k} is occupied while states \textbf{q} must be empty. The Feynman diagrams depict the photon absorption and emission processes with spin angular momentum conservation.}
\label{Fermi_golden_rule}
\end{figure}

We then focus on and decompose
the Zeeman coupling $\hat{V}(t)=\hat{V}^{(+)}e^{-i\omega t}+\hat{V}%
^{(-)}e^{i\omega t}$, where $\hat{V}^{(\pm)}=\sum_{\mathbf{k},\mathbf{k}
^{\prime}}\sum_{\eta,\epsilon}{\cal G}^{(\pm)}_{\mathbf{k}^{\prime
}-\mathbf{k}}|_{\eta\epsilon}e^{\mp i\omega t}|\mathbf{k}^{\prime},\eta\rangle\langle
\mathbf{k},\epsilon|$ with ${\cal G}^{(\pm
)}_{\mathbf{q}}|_{\eta\epsilon}=(\mu_{0}\gamma_{e}\hbar/2)\sum_{\alpha=\{x,y,z\}}\sigma
_{\eta\epsilon}^{\alpha}h_{\alpha}^{(\pm)}(\mathbf{q})$. On substituting Eq.~(\ref{polarizedfield}), only  ${\cal G}^{(+)}_{\bf q}|_{\uparrow
\downarrow}$ and ${\cal G}^{(-)}_{\bf q}|_{\downarrow\uparrow}$ are non-zero, reflecting the angular-momentum conservation.
The localized or focused magnetic field coherently couples electron states of different wave
vectors since the operator $\hat{V}^{(+)}$ has finite matrix elements between
an occupied initial state $|\mathbf{k},\uparrow\rangle$ of energy
$\varepsilon_{\mathbf{k}}$ and an empty state $|\mathbf{k}%
^{\prime},\downarrow\rangle$ of higher energy $\varepsilon_{\mathbf{k^{\prime}}}$.  Inversely, at finite temperatures
$\hat{V}^{(-)}$ emits a photon when the higher energy state is occupied and
the lower empty. This spin injection driven by the spin-flip induced by the AC magnetic field
\eqref{polarizedfield} in Eq.~(\ref{spin_injection_rate}) that may be understood in terms of photon absorption processes in Fig.~\ref{Fermi_golden_rule} for a parabolic
electron dispersion $\varepsilon_{\bf k}={\hbar^{2}k^{2}}/(2m^{\ast})$. The red curves sketch an
electron with spin \textquotedblleft$\downarrow$\textquotedblright\ that under
absorption of a photon with energy $\hbar\omega$ flips to a \textquotedblleft%
$\uparrow$\textquotedblright\ under energy and momentum conservation, i.e., a
transition from $|\mathbf{k},\downarrow\rangle$ to $|\mathbf{q}_{+}%
,\uparrow\rangle$. Here $\varepsilon(\mathbf{q}_{\pm})=\varepsilon
(\mathbf{k})\pm\hbar\omega$. The blue curves indicate the photon emission
process from $|\mathbf{k},\uparrow\rangle$ to $|\mathbf{q}_{-},\downarrow
\rangle$.

At zero temperature electrons occupy states below the Fermi energy $E_{f}
$ and Fermi wave number $k_{f}$. Substituting the parabolic electron
dispersion and the field \eqref{polarizedfield} into \eqref{spin_injection_rate},
\begin{align}
\frac{\partial\mathbf{s}}{\partial t}\Big|_{\mathrm{DC}} &
=\sum_{\mathbf{k}}\int_{0}^{2\pi}d\varphi_{\mathbf{q}}\int_{0}^{\infty
}d\varepsilon_{\bf q}\frac{iSm^{\ast}}{\pi\hbar^{2}}\left(  f(\varepsilon
_{\mathbf{q}})-f(\varepsilon_{\mathbf{k}})\right) \nonumber\\
&  \times\left(  \mathbf{h}_{\mathbf{k-q}}^{(+)\ast}\times\mathbf{h}%
_{\mathbf{k-q}}^{(+)}\right)  \delta\left(  \varepsilon_{\mathbf{k}%
}+\hbar\omega-\varepsilon_{\mathbf{q}}\right)+{\rm H.c.} \nonumber\\
&  ={\omega(m^{\ast}\mu_{0}\gamma_{e}h_{0})^{2}}/{(2\pi\hbar)}\hat{\bf z}.
\label{Fermi_golden}
\end{align}
It indicates that the photon absorption
in Fig.~\ref{Fermi_golden_rule} contributes to the spin injection. The number of absorbed photons
with frequency $\hbar\omega$ is proportional to $\hbar\omega D_{\mathrm{2DEG}%
},$ where the density of states of 2DEG $D_{\mathrm{2DEG}}=Sm^{\ast}%
/(2\pi\hbar^{2})$. In the ballistic regime, the energy and angular momentum injection rate into the electron system equals the total spin and energy currents flowing through a
circle around the source. We then estimate the spin current density by~\cite{supplement}
\begin{align}
\frac{\partial\mathbf{s}}{\partial t}\Big|_{\mathrm{DC}} &
=\varoiint\pmb{\cal J}_{s}^{\mathrm{DC}}(\pmb{\rho})\cdot d\mathbf{S}%
\nonumber\\
&  =2\pi\rho\mathcal{J}_{s}^{\mathrm{est}}({\rho}){\hat{\mathbf{z}}}%
=\omega(m^{\ast}\mu_{0}\gamma_{e}h_{0})^{2}/(2\pi\hbar){\hat{\mathbf{z}}%
},
\label{estimation}%
\end{align}
where $\rho=|\pmb{\rho}|$. The process is proportional to the absorption
coefficient of the light intensity $\sim h_{0}^{2}$.

\textit{Quantum formalism.}---Below we substantiate the magnetic spin pumping
found by rate equation as sketched above by a full quantum formalism. The Hamiltonian of 2DEG in the $x$-$y$ plane subject to an inhomogeneous AC
magnetic field $\mathbf{h}({\pmb\rho},t)$ of frequency $\omega$ reads
\begin{align}
\hat{H}_{e} &  =\hat{H}_{0}+\hat{V}(t)=\sum_{\mathbf{k,\eta}}\left(
\varepsilon_{\mathbf{k}}-\mu\right)  |\mathbf{k},\eta\rangle\langle
\mathbf{k},\eta|\nonumber\\
&  +\sum_{\zeta=\pm}\sum_{\mathbf{k},\mathbf{k}^{\prime}}\sum_{\eta,\epsilon=\{\uparrow,\downarrow\}
}{\cal G}^{(\zeta)}_{\mathbf{k}^{\prime}-\mathbf{k}}|_{\eta\epsilon}e^{-i{\zeta}\omega
t}|\mathbf{k}^{\prime},\eta\rangle\langle\mathbf{k},\epsilon|,
\label{H2DEG}
\end{align}
where all symbols have been defined above. Referring to the
SM~\cite{supplement} for details, the time evolution operator $\hat{U}%
_{I}(t,t_{0})$ in the interaction representation is expanded into Dyson series
of which we again retain the lowest two order of $\hat{V}$. The electron
wavefunction evolves under the perturbation according to $\psi_{\eta}(\pmb{\rho})=\sum_{\mathbf{k^{\prime
}}}\sum_{\epsilon}\langle\boldsymbol{\rho}|e^{-i\hat{H}_{0}t/\hbar}|\mathbf{k}%
^{\prime},\epsilon\rangle\langle\mathbf{k^{\prime}},\epsilon|\hat{U}%
_{I}(t,t_{0}\rightarrow-\infty)|\mathbf{k},\eta\rangle$~\cite{Sakurai,Mahan,Vignale}. The field operator
 of such driven eigenstates in terms of
the electron annihilation operator $\hat{a}_{\eta}(\mathbf{k})$ of an
unperturbed state with wave vector $\mathbf{k}$ and spin $\eta$~\cite{supplement}
\begin{align}
&\hat{\psi}_{\eta}(\boldsymbol{\rho},t)=\frac{1}{\sqrt{S}}\left(
\sum_{\mathbf{k}}\hat{a}_{\eta}(\mathbf{k})e^{i(\mathbf{k}\cdot
\boldsymbol{\rho}-\varepsilon_{\mathbf{k}}t/\hbar)}
+\sum_{\zeta=\pm}\sum_{\epsilon}
\sum_{\mathbf{k,k^{\prime}}}\hat{a}_{\epsilon}(\mathbf{k})\right. 
\nonumber\\
&\times\frac{{\cal G}_{\mathbf{k^{\prime}-k}}^{(\zeta)}|_{\eta\epsilon}e^{i(\mathbf{k^{\prime}}\cdot\boldsymbol{\rho}-(\varepsilon_{\mathbf{k}%
}+\zeta\hbar\omega)t/\hbar)}}{\varepsilon_{\mathbf{k}}+\zeta\hbar
\omega-\varepsilon_{\mathbf{k^{\prime}}}+i0_{+}}+\sum_{\zeta_{1},\zeta_{2}%
=\pm}\sum_{\mathbf{k^{\prime}},\mathbf{k},\mathbf{q}}\sum_{\xi,\epsilon=\uparrow,\downarrow}\hat
{a}_{\epsilon}(\mathbf{k})\nonumber\\
&  \left.  \times\frac{{\cal G}_{\mathbf{k}^{\prime
}-\mathbf{q}}^{(\zeta_{1})}|_{\eta\xi}{\cal G}_{\mathbf{q}-\mathbf{k}}^{(\zeta_{2})}|_{\xi\epsilon}e^{i(\mathbf{k^{\prime}}\cdot\boldsymbol{\rho}-(\varepsilon_{\mathbf{k}%
}+(\zeta_{1}+\zeta_{2})\hbar\omega)t/\hbar)}}{\left(  \varepsilon_{\mathbf{k}%
}+(\zeta_{1}+\zeta_{2})\hbar\omega-\varepsilon_{\mathbf{k^{\prime}}}%
+i0_{+}\right)  \left(  \varepsilon_{\mathbf{k}}+\zeta_{2}\hbar\omega
-\varepsilon_{\mathbf{q}}+i0_{+}\right)  }\right),
\nonumber
\end{align}
where the Cartesian position (wave) vector $\pmb{\rho}$ ${(\mathbf{q}}_{\zeta
})$ transforms to polar coordinates as $\pmb{\rho}=\rho\cos\varphi
\hat{\mathbf{x}}+\rho\sin\varphi\hat{\mathbf{y}}$ ($\mathbf{q}_{\zeta
}=q_{\zeta}\cos\varphi_{q_{\zeta}}\hat{\mathbf{x}}+q_{\zeta}\sin
\varphi_{q_{\zeta}}\hat{\mathbf{y}}$).

The spin-current density carried by the excited states 
\[
\pmb{\cal J}_{s}(\boldsymbol{\rho},t)=\left\langle \frac{\hbar^{2}}{4im^{\ast
}}\sum_{\eta\epsilon}\hat{\psi}_{\eta}^{\dagger}{\pmb\sigma}_{\eta\epsilon
}\nabla\hat{\psi}_{\epsilon}+\mathrm{H.c.}\right\rangle =\sum_{n\geq
0}\pmb{\cal J}_{s}^{(n)}(\boldsymbol{\rho},t),
\]
where the second step indicates a perturbation expansion $\pmb{\cal J}_{s}%
^{(n)}\propto V^{n}$. Since the ensemble average $\left\langle \hat{a}_{\eta
}^{\dagger}(\mathbf{k}_{1})\hat{a}_{\epsilon}(\mathbf{k}_{2})\right\rangle
=\delta_{\mathbf{k}_{1}\mathbf{k}_{2}}\delta_{\eta\epsilon}f_{\mathbf{k}}$ in
terms of the Fermi-Dirac distribution $f_{\mathbf{k}}$, the
zero-order (equilibrium) spin current
$\pmb{\cal J}_{s}^{(0)}(\boldsymbol{\rho},t)=\hbar^{2}/({4m^{\ast}S})\sum_{\mathbf{k}}f_{\mathbf{k}}\Tr\left(  \pmb{\sigma}\otimes\mathbf{k}%
\right)  +\mathrm{H.c.}$
vanishes. 
In the linear response, the tensor
\begin{align}
\pmb{\cal J}_{s}^{(1)}(\pmb{\rho})&=\sum_{\mathbf{k}}\frac
{-i}{8\pi S}\int_{\varphi-\frac{\pi}{2}}^{\varphi+\frac{\pi}{2}%
}d\varphi_{q_{+}}(f_{\mathbf{k}}-f_{\mathbf{q}_{+}})e^{i((\mathbf{q}_{+}-\mathbf{k})\cdot{\pmb{\rho}}%
-\omega t)}\nonumber\\
&  \times\Tr\left(  \mathcal{G}_{\mathbf{q}_{+}-\mathbf{k}}^{(+
)\dagger}\pmb{\sigma}\otimes(\mathbf{k}+\mathbf{q}_{+})\right)+\mathrm{H.c.}
\nonumber
\end{align}
oscillates at frequency $\omega$ and can be detected via the AC spin Hall
effect~\cite{HuJunJiao,DahaiWei}. However, the linear response conserves
energy and spin and vanishes on time average. Analogous to the spin pumping by
magnetization dynamics, DC spin currents emerge in the second-order term of
the perturbation series:
\begin{align}
&  \pmb{\cal J}_{s}^{(2)}(\boldsymbol{\rho},t)=\sum_{\zeta_{1},\zeta_{2}=\pm}
\sum_{\mathbf{q},\mathbf{q}^{\prime},\mathbf{k}}\frac{\hbar^{2}}{4m^{\ast}%
}\mathrm{Tr}\left(  \mathcal{G}_{\mathbf{q}-\mathbf{k}}^{(\zeta_{1})\dagger
}(\pmb{\sigma}\otimes\mathbf{q}^{\prime})\mathcal{G}_{\mathbf{q}^{\prime
}-\mathbf{k}}^{(\zeta_{2})}\right) \nonumber\\
&  \times\left[  \frac{f_{\mathbf{k}}}{(\varepsilon_{\mathbf{k}}+\zeta
_{2}\hbar\omega-\varepsilon_{\mathbf{q^{\prime}}}+i0_{+})(\varepsilon
_{\mathbf{k}}+\zeta_{1}\hbar\omega-\varepsilon_{\mathbf{q}}-i0_{+})}\right.
\nonumber\\
&  +\frac{f_{\mathbf{q}^{\prime}}}{(\varepsilon_{\mathbf{q}}+(\zeta_{2}%
-\zeta_{1})\hbar\omega-\varepsilon_{\mathbf{q^{\prime}}}+i0_{+})(\varepsilon
_{\mathbf{k}}+\zeta_{2}\hbar\omega-\varepsilon_{\mathbf{q^{\prime}}}+i0_{+}%
)}\nonumber\\
&  \left.  +\frac{f_{\mathbf{q}}}{(\varepsilon_{\mathbf{q}}+(\zeta_{2}%
-\zeta_{1})\hbar\omega-\varepsilon_{\mathbf{q^{\prime}}}+i0_{+})(\varepsilon
_{\mathbf{q}}-\zeta_{1}\hbar\omega-\varepsilon_{\mathbf{k}}+i0_{+})}\right]
\nonumber\\
&  \times e^{i((\mathbf{q^{\prime}}-\mathbf{q})\cdot\pmb{\rho}-(\zeta
_{2}-\zeta_{1})\omega t)}+\mathrm{H.c.},
\nonumber
\end{align}
in which the $\zeta_{1}=\zeta_{2}$ contribution is constant in time and leads
to the DC spin current
\begin{align}
\pmb{\cal J}_{s}^{\mathrm{DC}}(\boldsymbol{\rho}) &  \approx\sum_{\zeta=\pm}\sum
_{\mathbf{k}}i\frac{m^{\ast}\mu_{0}^{2}\gamma_{e}^{2}}{32\pi^{2}S}%
\int_{\varphi-\frac{\pi}{2}}^{\varphi+\frac{\pi}{2}}d\varphi_{p_{\zeta}%
}d\varphi_{q_{\zeta}}e^{i(\mathbf{q}_{\zeta}-\mathbf{p}_{\zeta})\cdot
\boldsymbol{\rho}}\nonumber\\
&  \times\left(  \mathbf{h}^{(\zeta)}(\mathbf{q}_{\zeta}-\mathbf{k}%
)\times\mathbf{h}^{(\zeta)\ast}(\mathbf{p}_{\zeta}-\mathbf{k})\right)
\otimes\mathbf{q}_{\zeta}\nonumber\\
&  \times(f_{\mathbf{k}}-f_{\mathbf{q}_{\zeta}})+\mathrm{H.c.},
\label{J^DC}
\end{align}
in the approximation $f_{\mathbf{q}}=f_{\mathbf{q^{\prime}}}$ due to the
factor $1/(\varepsilon_{\mathbf{q}}-\varepsilon_{\mathbf{q^{\prime}}}+i0_{+}%
)$. We derive the same result by the density-matrix approach in the
SM~\cite{supplement}. The circular polarization $\mathbf{h}^{(+
)}(\mathbf{q}_{+}-\mathbf{k})\times\mathbf{h}^{(+)\ast}(\mathbf{p}%
_{+}-\mathbf{k})$ or \textquotedblleft photon spin"~\cite{Bliokh,Lloyd} governs the electron
spin polarization, which depends on the optical/microwave source and is flexibly tunable.

The above formalism holds for arbitrary magnetic field profiles, frequencies,
and electron densities. It is convenient to derive specific results from a line
source $\mathbf{h}(\boldsymbol{\rho},t)=\mathbf{h}(x,t)$ with Fourier
components $\mathbf{h}^{(\zeta)}(\mathbf{q})=2\pi\delta(q_{y})\mathbf{H}%
^{(\zeta)}(q_{x})$. In the far-field $x\rightarrow+\infty$, the DC spin
current~\cite{evanescent_pumping}
\begin{align}
&\pmb{\cal J}^{\mathrm{DC}}_{s,{\rm 1D}}(\boldsymbol{\rho})  \approx\sum_{\zeta=\pm}
\sum_{k_{x},k_{y}}\frac{im^{*}\mu_{0}^{2}\gamma_{e}^{2} }{8\kappa_{\zeta}%
S}\left( f(k_{x},k_{y})-f(\kappa_{\zeta},k_{y})\right) \nonumber\\
& \times\left( \mathbf{H}^{(\zeta)}(\kappa_{\zeta}-k_{x})\times\mathbf{H}%
^{(\zeta)*}(\kappa_{\zeta}-k_{x})\right) \otimes\hat{\mathbf{x}}+\mathrm{H.c.},
\nonumber
\end{align}
where $\kappa_\zeta\equiv\sqrt{k_{x}^{2}+2\zeta m^{*} \omega/\hbar}$.

\textit{Numerical results}.---Here we illustrate the 2D spin radiation by the THz field \eqref{polarizedfield} of strong localization. Substituting into Eq.~(\ref{J^DC}),
\begin{align}
\pmb{\cal J}^{\mathrm{DC}}_{s}(\boldsymbol{\rho})  & =\sum_{\mathbf{k}
}\frac{m^{*}\mu_{0}^{2}\gamma_{e}^{2}h_{0}^{2}}{8S}  \left(
f_{\mathbf{k}}-f_{\mathbf{q}_{+}}\right) \left( \hat{\mathbf{z}}%
\otimes\hat{\mathbf{e}}_{\rho}\right) \nonumber\\
&\times\left(kF(k\rho)+q_{+}F(q_{+}\rho)\right)     \nonumber\\
& \approx\frac{m^{*2}\mu_{0}^{2}\gamma_{e}^{2}h_{0}^{2}}{8\pi\hbar}\omega
k_fk_f F(k_f\rho)\left( \hat{\mathbf{z}}\otimes\hat{\mathbf{e}}_{\rho}\right),
\label{J_delta}%
\end{align}
where $q_+=|\mathbf{q}_+|$ and $J_{n}(x)$ and $H_{n}(x)$ are the $n$-order Bessel function of the first kind and Struve
function in $F(x)=J_{0}(x)H_{-1}(x)+J_{1}(x)H_{0}(x)$. In the second step, we assume the degenerate limit, in which only electrons near the Fermi surface contribute.

\begin{figure}[tbh]
\centering
\includegraphics[width=0.235\textwidth,trim=0.2cm 0.3cm 0.4cm
0.2cm]{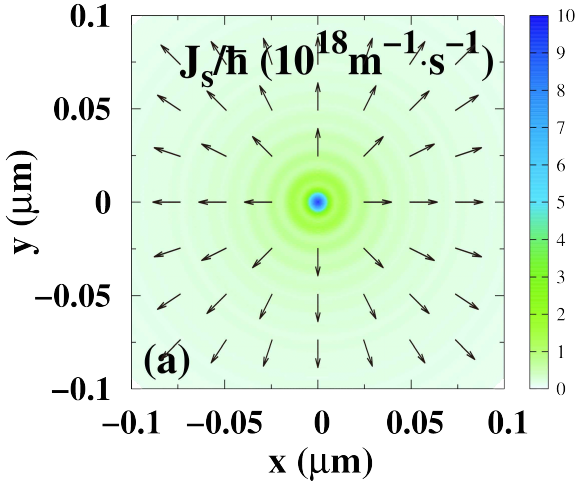} \hspace{0.25cm}
\includegraphics[width=0.206\textwidth,trim=0.5cm 0.3cm 0.6cm 1cm]{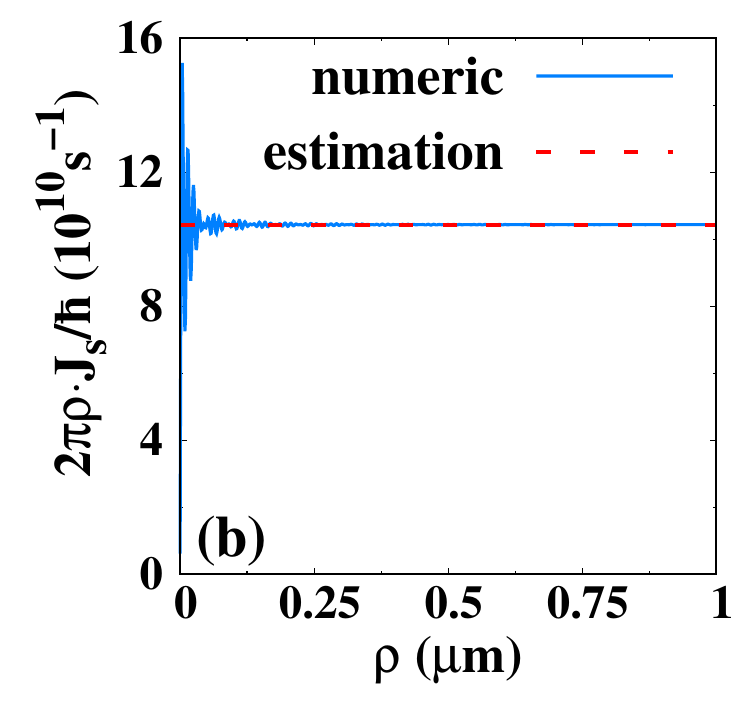}
\hspace{0.25cm}
\includegraphics[width=0.22\textwidth,trim=0.5cm 0.2cm 0.5cm 0cm]{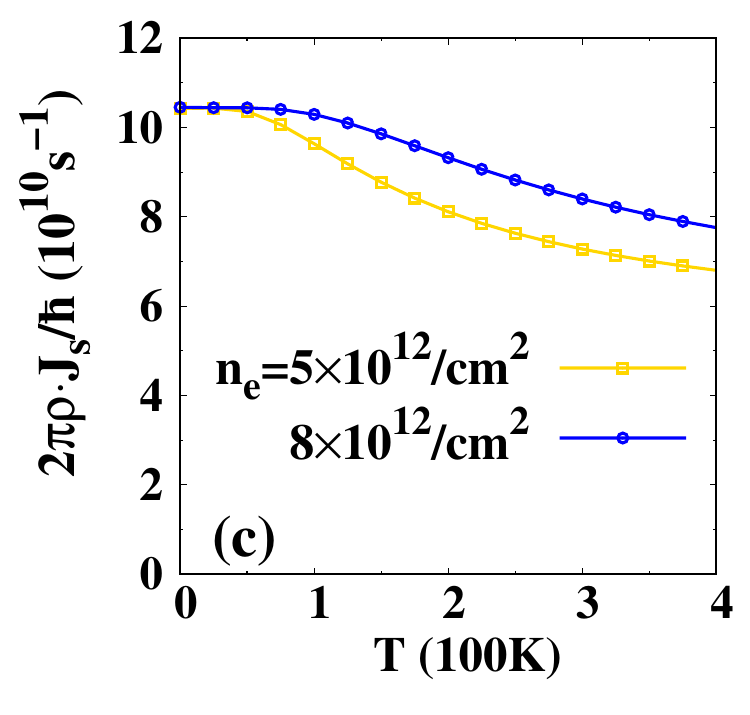}
\hspace{0.22cm} \includegraphics[width=0.22\textwidth,trim=0cm 0.2cm 0.9cm 0cm]{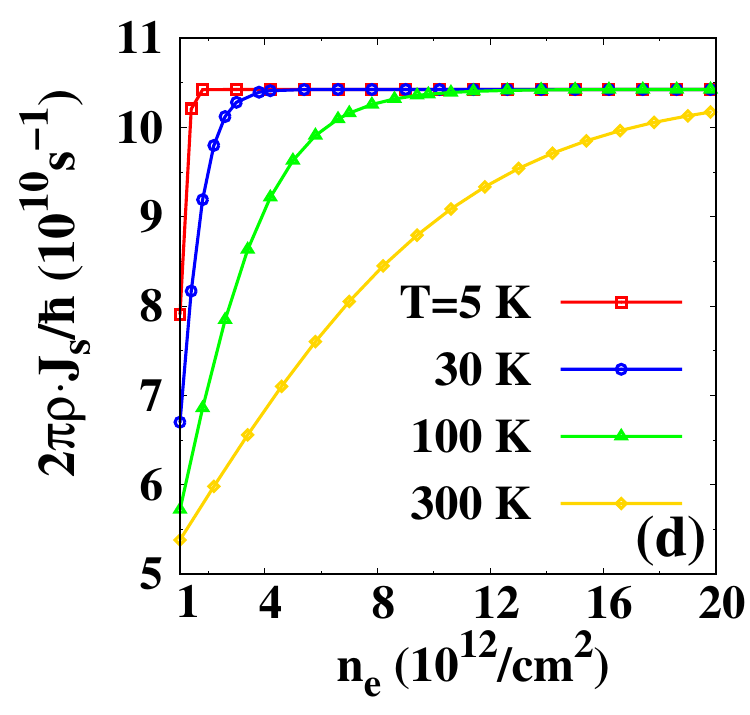}
\caption{Radiated DC spin current $\pmb{\cal J}^{\mathrm{DC}}_{s}(\pmb{\rho})$ under a spot of magnetic field. (a)
illustrates the magnitude (the color) and direction (the arrows) of the radiated spin current. (b) plots $2\pi\rho{\pmb{\cal J}}^{\mathrm{DC}}_{s}(\rho)/\hbar$ as a function of radius $\rho$ and compares with the estimation of spin transfer rate via Fermi's golden rule [Eq.~\eqref{estimation}]. (c) and (d) plot the efficiency of the spin transfer with different temperatures and electron densities.}
\label{Js_delta}
\end{figure}

Figure~\ref{Js_delta} plots the spatial distribution of radiated $\pmb{\cal J}_{s}^{\mathrm{DC}}(\boldsymbol{\rho})$ by
considering a monolayer $n$-doped $\mathrm{MoS_{2}}$ with effective electron mass
$m^*=0.48m_{e}$~\cite{MoS2_mass} and $g$-factor $|g_{e}%
|=2.16$~\cite{Marinov}. A typical electron density $n_{e}=5.6\times10^{12}%
$~$\mathrm{cm}^{-2}$~\cite{Siao} corresponds to a Fermi energy $E_{f}\sim
28.5$~meV. The electron mobility is high~\cite{Radisavljevic} and the lifetime of out-of-plane spin polarization is long ~\cite{LWang}. The field frequency $\omega=10$~THz and its amplitude $\mu_{0}h_{0}\approx\pi\times10^{-16}$~$\mathrm{T}\cdot\mathrm{m}^{2}$ is equivalent to a spot of magnetic field $0.4$~mT of radius
$500$~nm, which could be generated by THz near-field from metallic nanoparticles~\cite{nano_optics,plasmonics} or SNOM~\cite{Betzig,LeWang,HaominWang,Vincent}. As shown in Fig.~\ref{Js_delta}%
(a), the spin current radiates outward from the optical spot and decays
according to $1/\rho$. In Fig.~\ref{Js_delta}(b), $2\pi\rho{\pmb {\cal J}}_{s}^{\mathrm{DC}}(\rho)$ is almost a constant with the increase of radius
$\rho$, which agrees well with the estimation from the spin transfer in
Eq.~(\ref{estimation}). The effect is robust and persisting at different
temperatures and electron densities, as shown in Fig.~\ref{Js_delta}(c) and (d). 

\textcolor{blue}{The efficiency of spin-current generation by optical pumping competes with that proposed in Ref.~\cite{HongyiYu} by band anisotropy, although they may not appear simultaneously. According to Ref.~\cite{HongyiYu}, the electric field of magnitude $E$ drives the spin current $|{\bf J}_s|/\hbar\sim (12\pi/\hbar)E_f|\beta| (e E\tau/\hbar)^2$, where $\beta$ characterizes the band anisotropy and $\tau$ is the momentum relaxation time. By the mobility $\mu_e=10^3$~${\rm cm}^2~{\rm V}^{-1}~{\rm s}^{-1}$~\cite{Baugher}, $\tau=\mu_e m^*/e\sim 0.27$~ps at low temperature. Using $\beta=-0.49$~\AA~ and $E=10$~mV~$\mu{\rm m}^{-1}$ in Ref.~[6], 
the spin current $|{\bf J}_s|/\hbar\sim 10^{18}$~${\rm m}^{-1}$~${\rm s}^{-1}$ when $E_f=28.5$~meV is not larger than $|{\bf J}_s|/\hbar\sim 10^{19}$~${\rm m}^{-1}$~${\rm s}^{-1}$ by optical pumping.}

\textcolor{blue}{Recent works demonstrated that the ferromagnetic resonance of nanomagnets can generate local circularly polarized magnetic fields, which have been used to pump magnons in the nearby thin magnetic films~\cite{Baumgaertl,evanescent_pumping,HanchenWang,2D_nonhermitian,Kamenetskii}.} The optical spin pumping of conductors may also be realized by focused optical fields of a small radius~\cite{nano_optics,plasmonics,Betzig,LeWang,Vincent,HaominWang}. Referring to the SM~\cite{supplement}, the spin current can be efficiently generated by realistic magnetic-field spots and for the spin radiation by a line source, the directed spin current does not decrease with the distance.

\textit{Optical radiation to exciton spin/valley current}.---We then generalize the above mechanism to the spin-current generation of charge-neutral angular
momentum-carriers such as excitons and chiral phonons~\cite{phonon_spin_1,phonon_spin_2,phonon_spin_3}. The energy-degenerate excitons in opposite valleys of monolayer MoS$_{2}$ carry
opposite spins $\{\uparrow,\downarrow\}$. Their coupling with magnetic
fields is referred to as ``valley Zeeman
effect"~\cite{valley_Zeeman_1,valley_Zeeman_2,valley_Zeeman_3}. Exciton with a valley, thereby spin, polarization can be pumped by a laser of circular polarization via direct-band photon absorption. We
demonstrate here that an exciton pure spin or valley current radiates when the exciton distribution is subjected to a focused THz magnetic field, differing from previous proposals based on interference~\cite{Asgari,Sharma} or dispersion warping effect~\cite{WenYuShan}.

\begin{figure}[tbh]
\centering
\includegraphics[width=0.22\textwidth,trim=1cm 0.5cm 1.5cm 0.7cm]{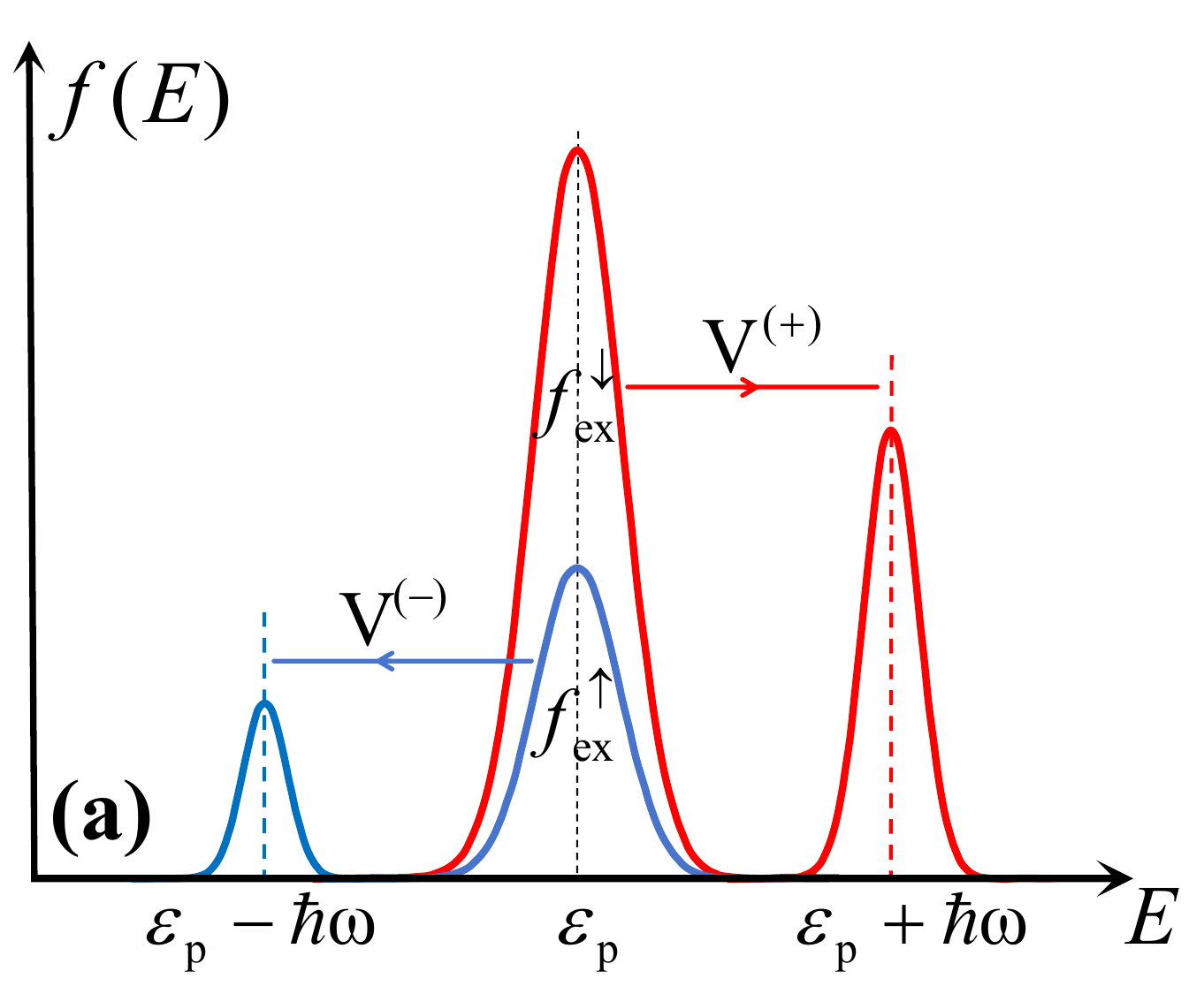} \includegraphics[width=0.22\textwidth,trim=0.4cm 1.3cm 1cm 0cm]{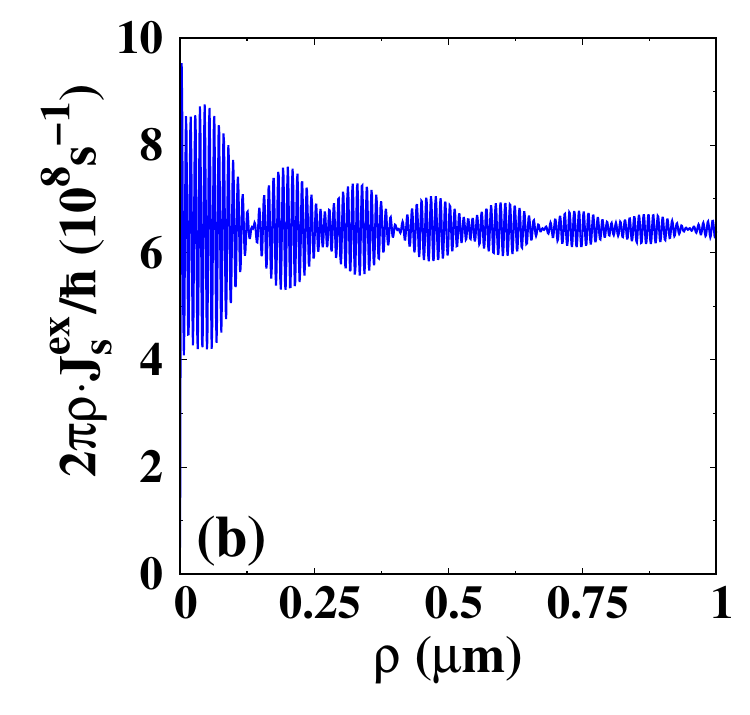}
\caption{Optical radiation of exciton spin
current. (a) addresses the spin transfer process in the opposite valleys, in which exciton distributions $f^{\uparrow}_{\rm ex}\ne f^{\downarrow}_{\rm ex}$. (b) shows pumped exciton spin current $2\pi\rho
\pmb{\cal J}^{\rm ex}_{s}(\rho)/\hbar$. }%
\label{Js_exciton}
\end{figure}

Here an optical laser generates a Gaussian distribution   $f_{\mathrm{ex}}^{\uparrow,\downarrow
}(\mathbf{k})=\alpha_{\rm ex}^{\uparrow,\downarrow}\exp(-(\varepsilon_{\mathbf{k}%
}^{\mathrm{ex}}-\varepsilon_{p})^{2}/(  2\delta_{\varepsilon}^{2})
)$ of excitons with spin $\{\uparrow,\downarrow\}$ in opposite valleys, centered around the laser energy $\varepsilon_{p}$ and broadened by $\delta_{\varepsilon}$~\cite{exciton_Yu,CMai,QWang}. $\varepsilon
_{\mathbf{k}}^{\mathrm{ex}}=\hbar^{2}k^{2}/(2m_{\mathrm{ex}}^{\ast})$ with exciton mass $m_{\mathrm{ex}}^{\ast}$. The normalization factor
$\alpha_{\rm ex}^{\uparrow,\downarrow}={n_{\mathrm{ex}}^{\uparrow,\downarrow}%
}/{\sum_{\mathbf{k}}\exp[-(\varepsilon_{\mathbf{k}}^{\mathrm{ex}}%
-\varepsilon_{p})^{2}/\left(  2\delta_{\varepsilon}^{2}\right)  ]}$ depends on
the exciton densities $n_{\mathrm{ex}}^{\uparrow,\downarrow}$ in opposite 
valleys, tunable by the ellipticity of light polarization. We
then subject the exciton to a strongly localized magnetic field $\mathbf{h}(\boldsymbol{\rho},t)$ with circular polarization along the $\hat{\mathbf{z}}$-direction. For the excitons of spin $s$ and gyromagnetic
ratio $\gamma_{\mathrm{ex}}=g_{\mathrm{ex}}\mu_{B}$ with exciton $g$-factor $g_{\mathrm{ex}}$, the exciton spin current pumped by a focused magnetic field
reads, analogous to Eq.~(\ref{J_delta}),
\begin{align}
\pmb{\cal J}_{s}^{\mathrm{ex}}(\boldsymbol{\rho}) &  =\sum_{\mathbf{k}%
}\frac{m_{\mathrm{ex}}^{\ast}\mu_{0}^{2}\gamma_{\mathrm{ex}}^{2}h_{0}^{2}s^{2}}{2S}\left(  f_{\mathrm{ex}}^{\downarrow}(\mathbf{k})-f_{\mathrm{ex}}^{\uparrow}(\mathbf{q}_{+})\right) \nonumber\\
&\times \left(kF(k\rho)+q_{+}F(q_{+}\rho)\right)
\left(  \hat{\mathbf{z}}\otimes\hat{\mathbf{e}%
}_{\rho}\right).
\label{spin_current_exciton}
\end{align}
This exciton spin current vanishes when
$n_{\mathrm{ex}}^{\uparrow}=n_{\mathrm{ex}}^{\downarrow}$, but exists in the
presence of valley polarization, which can be understood from the pumping
process depicted in Fig.~\ref{Js_exciton}(a): The exciton with polarization
\textquotedblleft$\downarrow$" is pumped to the \textquotedblleft$\uparrow$"
states through the photon absorption $V^{(+)}$ process, and inversely, the
exciton with polarization \textquotedblleft$\uparrow$" is driven to the
\textquotedblleft$\downarrow$" states by photon emission $V^{(-)}$ process.
The net spin transfer by absorption and emission then depends on the exciton
distribution $f^{\uparrow}_{\rm ex}$ and $f^{\downarrow}_{\rm ex}$. 

Figure~\ref{Js_exciton}(b) plots the exciton spin current under a laser
irradiation of energy $\varepsilon_{p}\approx100$~meV and bandwidth
$\delta_{\varepsilon}\approx 4$~meV~\cite{exciton_tau}, and for the THz field  $\omega=10$~THz and $\mu_{0}h_{0}\approx\pi\times10^{-16}$~$\mathrm{T}%
\cdot\mathrm{m}^{2}$. For excitons,
$m_{\mathrm{ex}}^{\ast}\approx0.19m_{e}$~\cite{exciton_mass}, $g_{\mathrm{ex}%
}=-4$~\cite{exciton_g},  and $n_{\mathrm{ex}}^{\uparrow}=0.7n_{\mathrm{ex}%
}^{\downarrow}=7\times10^{9}$~$\mathrm{cm}^{-2}$~\cite{exciton_density}. \textcolor{blue}{A practical approach to measuring the exciton spin current is measuring valley accumulation at the opposite edges of samples, at which the accumulated excitons come from the K and K’ valleys that emit the light of opposite circular polarization that can be picked up by spatial resolved photoluminescence~\cite{Onga,Lundt}.}

For phonon spin current, we only need to replace the distribution function of exciton \eqref{spin_current_exciton} with that of chiral phonons~\cite{phonon_spin_1,phonon_spin_2,phonon_spin_3}.  We will address such effects and those of
different materials in the future.

\textit{Conclusion and discussion}.---In conclusion, we generalize the spin-current generation in static Stern-Gerlach effect to a dynamic scenario or ``magnetic spin pumping", in which a focused AC magnetic field provides ``forces" to radiate the spin current of charge-neutral excitons and phonons, as well as electrons. This effect is free of charge, optical selection rules, and spin-orbit coupling.
Its efficiency is measurable: the pumped spin current is of the same order as the spin Hall current generated by an electric field of 0.1~kV/cm and a common spin Hall conductivity $\sigma_x^y=10^5~(\Omega\cdot{\rm m})$~\cite{Sinova,Jungwirth}.
The polarization of pumped spin currents is governed by the angular momentum of optical/microwave fields, which is thereby tunable and beyond that limited by the magnetization direction in the spin pumping and the spin-Hall conductivity tensor.  The spin radiation from electromagnetic fields, which can hold longitudinal, transverse, and orbital angular momenta, to distinct mobile carriers brings a unified and efficient paradigm in their spin transport, bridging different fundamental concepts in spintronics, nano-optics~\cite{optic1,optic2,Lodahl}, and plasmonics~\cite{Bliokh,plasmonic}.

\begin{acknowledgments}
This work is financially supported by the National Key Research and Development Program of China under Grant No.~2023YFA1406600, the National Natural Science Foundation of China under Grant No. 12374109, and the startup grant of Huazhong University of Science and Technology. We thank Gerrit Bauer for the useful discussions. 
\end{acknowledgments}


\begin{thebibliography}{99}
\bibitem {footnote}The local optical field can be equivalently expressed in
the form $\mathbf{h}({\pmb \rho},t)=\sum_{\mathbf{q}}\left( \mathbf{H}
^{*}(\mathbf{-q})e^{i\omega t}+\mathbf{H}(\mathbf{q})e^{-i\omega t}\right)
e^{i\mathbf{q}\cdot{\pmb \rho}}$.
\end{thebibliography}
\end{document}